\documentclass[aps,prl,twocolumn,superscriptaddress
,showpacs]{revtex4-1}
\usepackage{mathbbol,amsmath,amsfonts,amssymb,bm,color,dsfont}
\usepackage{graphicx,psfrag}
\usepackage{epstopdf}

\def\im{{\sf i}}
\def\r{{\bm{r}}}
\def\k{{\bm{k}}}
\def\x{{\bm{x}}}
\def\i{{\bm{e_1}}}
\def\j{{\bm{e_2}}}

\begin{document}

\title{Commensurate and Incommensurate States of Topological Quantum Matter}

\author{Ashley Milsted}
\affiliation{Institut f\"ur Theoretische Physik, Leibniz Universitat Hannover, 
Appelstrasse 2, 30167 Hannover, Germany}
\author{Emilio Cobanera}
\affiliation{Instituut-Lorentz, Universiteit Leiden, P.O. Box 9506, 2300
RA Leiden,  The Netherlands}
\email[Electronic address: ]{cobanera@lorentz.leidenuniv.nl}
\author{Michele Burrello}
\affiliation{Max-Planck-Institut f\"ur Quantenoptik, Hans-Kopfermann-Str. 1, 
D-85748 Garching, Germany}
\author{Gerardo Ortiz} 
\affiliation{Department of Physics, Indiana University, Bloomington,
IN 47405, USA}

\date{\today}

\begin{abstract}
We prove numerically and by dualities the existence of modulated,
commensurate and incommensurate states of topological quantum matter in simple systems 
of parafermions, motivated by recent proposals for the realization of such 
systems in mesoscopic arrays. In two space dimensions, we obtain the simplest 
representative of a topological universality class that we call Lifshitz. It is 
characterized by a topological tricritical point where a non-locally ordered 
homogeneous phase meets a disordered phase and a third phase that displays 
modulations of a non-local order parameter. 
\end{abstract}
\maketitle

In recent years, most efforts directed at investigating   
topological quantum matter experimentally have taken a top-to-bottom approach, 
starting from model Hamiltonians and engineering a systems to realize it.
From this point of view, mesoscopic superconducting arrays have already been 
proven successful \cite{ioffe09}, and also for cold atomic gases the implementation 
of topological phases of matter seems within reach \cite{goldman13}.

Inevitably, the model Hamiltonians in question can only be realized up to 
implementation-dependent modifications, that, although small, may be relevant in 
the sense of the renormalization group and drive large systems away from 
the intended topological phase. This practical aspect of the theory of phase 
transitions for topological quantum matter is the natural counterpart of 
analogous considerations for conventional systems like magnetic memories, 
which can only tolerate some range of temperatures and applied magnetic fields. 
However there is one crucial difference. Since a Landau theory of {\it non-local} 
order parameters, which are those appropriate to topological 
quantum matter, does not exist yet, it is difficult to predict and classify 
interacting topological gapless phases. By contrast, the classification of gapped 
phases is understood (for parafermions, see \cite{Bondesan2013,Motruk2013}).

In this paper we extend the list of demonstrated topological critical behaviors 
(see, for example, \cite{Ardonne2004,Feiguin2007,Tupitsyn2010,Dusuel2011,Schulz2012}). 
We will show that topological quantum matter can be driven into phases 
characterized by non-local orders incommensurate with the underlying lattice. 
Remarkably, it will become clear that modulated and floating (and, in particular,
incommensurate) topological quantum orders can easily arise in mesoscopic arrays 
from very natural interactions. And we will prove the existence 
of a topological universality class surprisingly sensitive to an underlying
lattice structure by locating a topological Lifshitz tricritical point
in the phase diagram of a two-dimensional model of topological quantum matter.

But let us recall first the basics of modulated Landau orders. 
When a {\it local} order parameter \(\Phi(\x)\) emerges in a lattice system, 
phases may occur in which this order parameter displays modulations 
\(\Phi(\x)\sim \Phi_0\cos(\k_0\cdot \x +\phi_0)\) commensurate with 
the lattice periodicity. The wave vector \(\k_0\) is restricted by the 
Lifshitz condition to take one of a few possible values in the first 
Brillouin zone \cite{Landau1968}. 

This picture of modulated local orders can break down  if interactions 
that favor competing periodicities are present, as exemplified by the ANNNI model
\cite{Bak1982,Selke1988} of magnetic ordering in the heavy lanthanoids \cite{Elliott1961}. 
In systems with such competing interactions, there might be regimes 
where the equilibrium wave vector varies continuously with some driving force, as 
first predicted in Ref. \cite{Hornreich1975} from the Landau functional density
\begin{eqnarray}\label{k_instability}
f=\kappa_2 \Phi^2+\kappa_4\Phi^4+\kappa_6\Phi^6+\gamma_1(\nabla \Phi)^2
+\gamma_2(\nabla^2\Phi)^2
\end{eqnarray}
for an Ising order parameter. Just as the standard Ising tricritical
point emerges at \(\kappa_2=\kappa_4=0\), the Lifshitz tricritical point 
emerges at \(\gamma_1=0\). It is the coexistence point for the paramagnetic, 
ferromagnetic, and modulated phases of the local order parameter \(\Phi\). 
On the coexistence line between the modulated 
and paramagnetic phases, starting at the Lifshitz point, 
the wave vector \(\k\) varies continuously with the driving field and so 
an additional critical exponent appears. If \(\k\) happens to vary 
continuously in a phase, then the phase is called floating.

In the following we demonstrate through explicit examples 
that the full range of phenomena associated with commensurate and 
incommensurate modulations and the Lifshitz point can also be present 
in topological quantum matter, but now in terms of non-local order parameters. 
Unlike the situation for local (Landau) orders just
discussed, there is no obvious way to predict such topological quantum orders 
on the basis of some general Landau-Wilson functional. 
This point showcases one of the troubling limitations in our current 
understanding of topological quantum matter at criticality.   

We start by considering a one-dimensional effective Hamiltonian with a discrete global 
$\mathbb{Z}_{2m}$ (\(m=1,3,\dots\) odd) symmetry that displays a 
critical floating phase. One may obtain a $\mathbb{Z}_{2m}$ symmetry in systems with
quasiparticles of fractional charge $e/m$ subjected to proximity-induced
superconducting pairing. The combination of these two ingredients provides
a channel for Cooper pairs to split into $2m$ indistinguishable parts.
Then the condensation of the Cooper pairs leads to a peculiar cyclic behavior 
of the local, charged degrees of freedom and induces the required 
$\mathbb{Z}_{2m}$ symmetry. These ideas are central to several 
proposals \cite{stern2012,shtengel2012,vaezi2013,cheng2013} that aim to realize 
localized parafermionic zero-energy modes (parafermions for short) in hybrid 
mesoscopic arrays including fractional topological insulators (FTI).
Parafermions are obtained by gapping the edge modes of a FTI and constitute a 
fractionalized version of Majorana zero-energy edge modes (Majoranas for short), 
allowing for the emergence of 1D systems which generalize \cite{fendley2012,
Bondesan2013,Motruk2013,Cobanera2014,Li2014} 
the well-known Majorana-Kitaev chain \cite{kitaev2001}.

\begin{figure}[t]
\includegraphics[width=8cm]{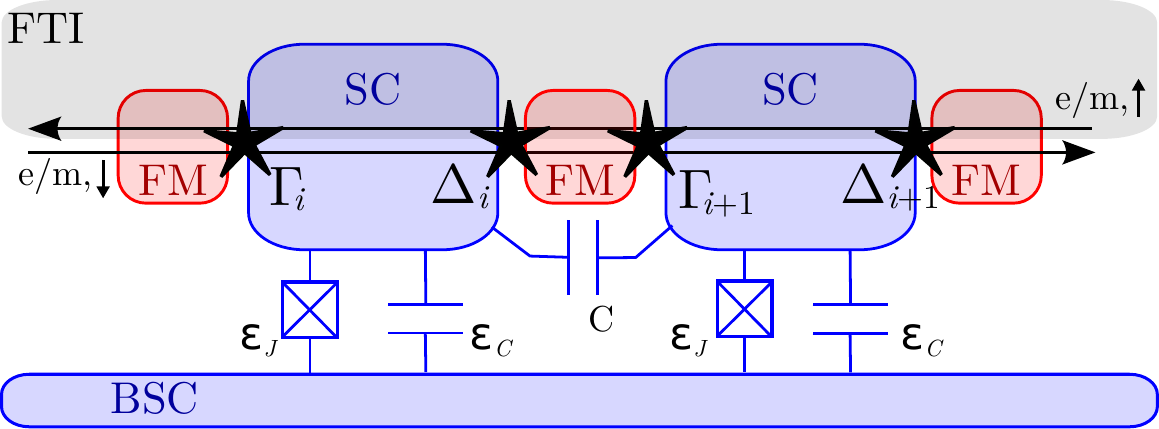}
\caption{Parafermions \(\Gamma_i\) and \(\Delta_i\) are localized
along the edge of a FTI at the interfaces between superconducting islands
(SC) and insulating ferromagnets (FM). Each superconducting island is coupled to
neighboring islands via a capacitive coupling $C$, and to a ground 
superconductor (BSC) via a Josephson junction $\varepsilon_J$ and a 
capacitive coupling $\varepsilon_C$.} 
\label{array_fig}
\end{figure}

Along the edge of an FTI, localized parafermions emerge at the 
interfaces between alternating regions where the edge modes of the FTI are 
gapped by proximity to superconducting islands or insulating ferromagnets 
\cite{stern2012,shtengel2012}, see Fig. \ref{array_fig}. 
Each superconducting island \(i\) hosts a pair of parafermionic modes 
\(\Gamma_i,\Delta_i\) sharing a fractional charge $q_i^{\sf f} = 0,\frac{1}{m},
\frac{2}{m}, \ldots,\frac{2m-1}{m}$, in units of $e$, defined modulo $2$ 
\cite{stern2012,shtengel2012}. 
Parafermions obey non-local commutation rules,
\begin{eqnarray}
\Gamma_i\Delta_j&=&e^{\im \frac{\pi}{m}} \, \Delta_j\Gamma_i\quad (i\leq j),\\
\Gamma_i\Gamma_j&=&e^{\im\frac{\pi}{m}} \, \Gamma_j\Gamma_i,\quad
\Delta_i\Delta_j=e^{\im \frac{\pi}{m}}\Delta_j\Delta_i \quad (i<j),\\
\Gamma_i^{2m}&=&\mathds{1}=-\Delta_i^{2m},\quad 
\Gamma_i^{\;}\Gamma_i^\dagger=\Delta_i^{\;}\Delta_i^\dagger=\mathds{1}.
\end{eqnarray}
This algebra of parafermions is a natural generalization of the Clifford algebra
of Majoranas. 

The charge $q_i^{\sf f}$ is the charge of the FTI edge segment 
coupled to the superconductor and may be represented by the 
operator $\Gamma_i^\dagger \Delta_i=e^{i\pi q_i^{\sf f}}$.
In our mesoscopic array, two main physical processes intervene to couple the 
zero-energy modes: a fractional Josephson effect \cite{shtengel2012,cheng2013}, 
which generalizes the electron tunneling mediated by Majorana modes \cite{xu2010},
and the charging interactions of the islands which, just like 
in the Majorana case \cite{vanheck2011,vanheck2012,hassler2012}, 
cause an energy splitting of the states with different fractional 
charges \cite{Burrello2013}. The Josephson interaction accounts for the tunneling 
of fractional quasiparticles between two neighboring islands and it changes their 
fermionic number by $\pm 1/m$. The tunneling of a single fractional charge is the 
dominant process and, in terms of parafermionic modes, it reads
\(-\left({E_J}/{2}\right)(\Gamma_{i+1}\Delta_i^\dagger  + H.c.).\)
The charging interactions are modelled by assuming that each island is coupled to 
a background superconductor by a strong normal Josephson junction and a capacitive 
contact, with magnitudes $\varepsilon_J$ and $\varepsilon_C$ respectively. 
See Fig. \ref{array_fig}. 

Besides the contribution coming from Cooper pairs, the total charge in each 
island includes the charge $q^{\sf ind}_i$ induced by the neighboring potentials, 
and the fractional charge $q_i^{\sf f}$ associated to the parafermions. The effect of 
these two contributions is especially important if $\varepsilon_J \gg \varepsilon_C$, 
that is, in the transmon regime \cite{girvin07}. 
In this regime the low energy physics can be described by semiclassically assuming
that the superconducting phase of the island is approximately pinned to the minima 
of the Josephson energy. Then the charging energy causes an effective interaction 
$-\Delta_C\cos\left(\pi(q^{\sf f}_i+q^{\sf ind}_i)\right)
$, where $\Delta_C$ depends on the ratio 
$\varepsilon_J/\varepsilon_C$ \cite{girvin07}, and the cosine dependence is due 
to the Aharonov-Casher effect associated with $2\pi$-phase slips in states with 
different charges $q^{\sf f}_i+q^{\sf ind}_i$ \cite{vanheck2011,hassler2012}. 
Following \cite{Burrello2013}, this interaction may be written as 
\(
-({1}/{2}) \left(E_{C}^{(1)}
\Gamma_i^\dagger \Delta_i + H.c. \right), 
\)
where \(E_{C}^{(1)}=\Delta_C e^{-\im \pi q^{\sf ind}}\) is, in general, complex. 
It is possible to tune \(q^{\sf ind}\), using voltage gates in the system, 
to take the values $0$ or $1$ and thus obtain a positive or negative 
single-island charging energy term.

A further charging term appears in the presence of a cross-capacitance 
$C$ between neighboring islands. This term originates from the simultaneous 
$2\pi$-phase slip of both islands \cite{hassler2012} and reads
\(-E^{(2)}_C\cos\left(\pi(q_i^{\sf f} + q_{i+1}^{\sf f}+ q^{\sf ind}_i 
+ q^{\sf ind}_{i+1})\right)\).
In particular, we impose that all the induced charges share a common
value $q^{\sf ind}$. By tuning $q^{\sf ind}$ to add a unit of charge to each island
($q^{\sf ind}_i \to q^{\sf ind}_i +1$), the relative sign between 
the coupling strengths $E^{(1)}_C$ and $E^{(2)}_C=|E_{C}^{(2)}|e^{-\im 2\pi q_{\sf ind}}$ 
may be controlled. This cross-capacitance interaction is translated into a 
four-parafermion operator and, combining all the previous terms, 
we obtain an effective Hamiltonian 
\begin{eqnarray}\label{dual_pANNNI}
H_{\sf eff}&=&-\frac{1}{2}\sum_{i=1}^L[(E_J 
\Gamma_{i+1} \Delta_i^\dagger \\
&+&E^{(1)}_C \Gamma_i^\dagger \Delta_i +E_C^{(2)} \Gamma_i^\dagger 
\Delta_i\Gamma_{i+1}^\dagger \Delta_{i+1})+H.c.] \nonumber
\end{eqnarray}
for the description of the array of Fig. \ref{array_fig} in its low-energy sector 
with periodic boundary conditions. In the following, we will take \(E^{(1)}_C=1\) and
\(E_C^{(2)}\leq0\). Then \(H_{\sf eff}\) is closely connected
to a generalization of the ANNNI model (corresponding to $m=1$) to 
any odd $m$ \cite{supplemental_material}.

\begin{figure}[ht]
\includegraphics[angle=0, width=\columnwidth]{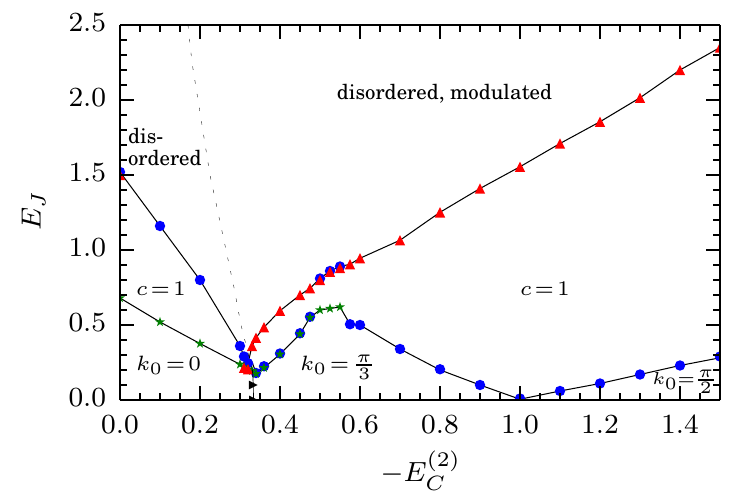}
\caption{Quantum phase diagram of \(H_{\sf eff}\) for $m=3$. The three ordered phases 
are labeled by the wave vector associated to the modulation of the string order 
parameter. The disordered phase contains modulated and unmodulated regimes, separated
by the (dashed) disorder line.  There are two critical phases with central charge $c=1$. 
Various indicators were used to mark the transitions: Blue dots mark
global maxima of the entanglement entropy, green stars mark local
maxima of its first $E_J$ derivative, red upwards triangles
mark local minima of the second $E_J$ derivative of the ground energy,
and black triangles mark discontinuities in the first derivative of the ground
energy.}
\label{fig_phase_diag}
\end{figure}

We studied the quantum phase diagram of \(H_{\sf eff}\) for \(m=3\) numerically,
computing approximate ground states using the open source evoMPS toolbox \cite{evoMPS}, 
which implements variational tangent plane techniques for matrix product states (MPS)
\cite{tangent_plane_mps}. In particular, evoMPS implements the nonlinear conjugate 
gradient method to accelerate the process significantly, particularly for critical regimes,
in comparison to imaginary time evolution \cite{Milsted2013}. We choose block translation 
invariant MPS with various block lengths in order to handle ground states with 
nontrivial periodicity.

The quantum phase diagram of \(H_{\sf eff}\) is shown in Fig. \ref{fig_phase_diag}.
There are three distinct gapped phases at low \(E_J\geq 0\) followed by  two 
critical phases, both with central charge \(c=1\) \cite{mps_cc}. The critical phases 
are topped by a gapped phase at large \(E_J\). To further characterize the 
(dis)orders in these phases, we follow the ideas of Refs. \cite{Cobanera2013,
vanHeck2014} to determine a non-local order parameter for \(H_{\sf eff}\) by mapping this
Hamiltonian to a Landau-ordered system. We obtain \cite{supplemental_material} that 
the ground-state \(|\Omega\rangle\) expectation value
\begin{equation}
\Sigma_i(d)=\langle \Omega|\prod_{n=i}^{i-d+1} 
\Gamma_n^\dagger\Delta_n^{\;}|\Omega\rangle
\end{equation} 
(independent of \(i\)) defines the required non-local 
order parameter. The string order parameter \(\Sigma_i(d)\) displays 
long-range order in the three gapped phases at small \(E_J\), with modulations 
characterized by \(k_0=0,\pi/3, \pi/2\). The wave vectors are ordered
as they appear for increasing \(-E^{(2)}_C\), see Fig. \ref{fig_phase_diag}. 
The ordered phases with \(k_0=0,\pi/3\) are separated by a first-order line.

\begin{figure}[t]
\includegraphics[width=8cm]{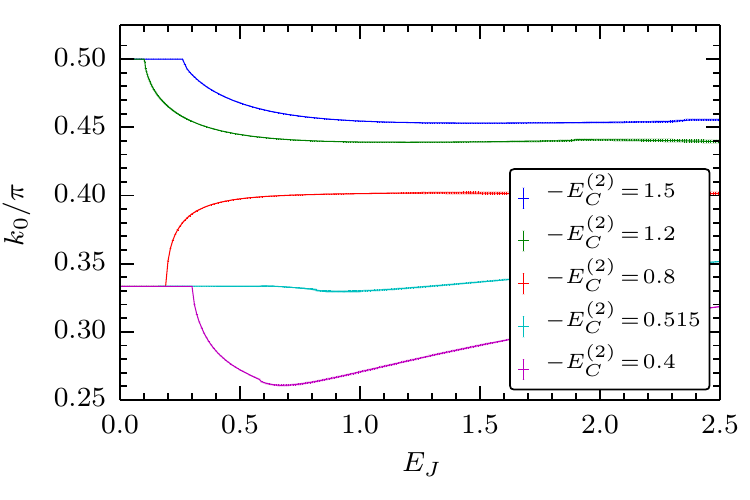}
\caption{Fitting of $\mathrm{Re}({\Sigma_i(d)})$ with a decay function
modulated with wave vector $k_0$. The wave vectors starts at a constant
value in the ordered phases and change continuously in 
the the critical phase on the right of the phase diagram. The error
bars represent statistical errors from the least-squares fitting.}
\label{variable_q}
\end{figure}

Starting at \(E_J = 0\) in either the gapped phase with \(k_0=\pi/3\) or \(k_0=\pi/2\)
and increasing \(E_J\) along a vertical line, 
the system enters the critical phase on the right in Fig. \ref{fig_phase_diag},
and the asymptotic behavior of \(\Sigma_i(d)\) changes from long-ranged
to algebraically decaying, but with a modulation \(k_0(E_J)\) that appears 
to vary continuously with \(E_J\) to the best available computer resolution.
In this regime, the periodicity of the non-local 
order in the system is no longer anchored to the lattice 
structure, and so our mesoscopic array demonstrates the existence of 
floating regimes for mesoscopically-realized topological quantum matter.
Fig. \ref{variable_q} shows \(k_0(E_J)\) for the full range of \(E_J\)
for three values of \(-E^{(2)}_C\) starting at \(k_0=\pi/3\),
and two values starting at \(k_0=\pi/2\).

As for the other phases, \(\Sigma_i(d)\) shows no modulations in the critical
phase on the left of the phase diagram. At \(-E^{(C)}_2=0\), this phase
is precisely \cite{supplemental_material} the critical phase of the \(p=2m=6\) 
clock model (see \cite{clock} and references therein). The modulations of the 
string order parameter survive in the the gapped, disordered phase at large \(E_J\) where  
\(\Sigma_i(d)\) decays exponentially fast in \(d\), but only for sufficiently 
large values of \(-E^{(2)}_C\). There is a regime in the disordered phase 
without modulations, as shown in Fig. \ref{fig_phase_diag}. The separation 
between the two disordered regimes, unmodulated and modulated, is called the 
disorder line in the literature on the ANNNI model. 

The string-ordered phases of \(H_{\sf eff}\) manifest the various ways in which 
the global, discrete symmetry
\begin{equation}
U_{\sf C}=\prod_{i=1}^L 
\Gamma_i^\dagger \Delta_i,\quad U_{\sf C}^{2m}=\mathds{1},
\quad \left[ U_{\sf C}, H_{\sf eff}\right] =0\,,
\end{equation}
can be spontaneously broken in the limit of infinite system size. 
There are however topological quantum orders that emerge
without spontaneously breaking any symmetries, as first noticed for Ising 
gauge theories \cite{Wegner1971}. These states of topologically quantum
matter are often modelled by systems with local symmetries, since,
by Elitzur's theorem \cite{Elitzur1975}, local symmetries cannot be spontaneously 
broken. The remainder of the paper focuses on a model that displays incommensurate 
behavior, and even a full-fledged topological Lifshitz point, without spontaneous 
symmetry breaking. We call a Lifshitz point topological if it is a tricritical 
point of the Lifshitz type, but associated to non-local orders only.   

The model in question, inspired by the mesoscopic realization 
of the toric code in terms of Majoranas \cite{Terhal2012}, 
features parafermions \(\Gamma_{(\r,\mu)},\Delta_{(\r,\mu)}\)
(\(\mu=1,2\)) on each link \((\r,\mu)\) connecting sites \(\r, \r+\bm{e_\mu}\)
of a square lattice. Let us define plaquette operators  \(
B_\r=U_{(\r,1)}U_{(\r+\i,2)}U_{(\r+\j,1)}^\dagger U_{(\r,2)}^\dagger
\) (in terms of the shorthand notation 
\(U_{(\r,\mu)}= \Gamma_{(\r,\mu)}^\dagger \Delta_{(\r,\mu)}\)),
and star operators
\( A_\r= 
\Delta_{(\r,1)}\Gamma_{(\r-\j,2)}^\dagger 
\Delta_{(\r,2)}\Gamma_{(\r-\i,1)}^\dagger.\)
As the naming suggests, the star and plaquette operators generate a commutative 
algebra. The gapped Hamiltonian  
\begin{eqnarray}\label{htc}
H_{\sf TC}=-\frac{1}{2}\sum_\r[h_p B_\r+h_s A_\r+H.c.]
\end{eqnarray}
is precisely the parafermionic representation of the \(\mathds{Z}_{2m}\)
toric code. In the following we will study the effect of the perturbation
\begin{equation}\label{Vpert}
V=-\frac{J_1}{2}\sum_{\r,\mu}[U_{(\r,\mu)}+H.c.]
-\frac{J_2}{2} \sum_\r [U_{(\r,2)}U_{(\r+\i,2)}+H.c.] 
\end{equation}
with \(J_1,-J_2\geq0\). Since the plaquette operators \(B_\r\) commute 
with the full Hamiltonian \(H_{\sf LTC}=H_{\sf TC}+V\), they
play the role of local symmetries. The ground state of the system belongs to the 
gauge-invariant sector where  the plaquettes \(B_\r\) act as the identity. 

For the purpose of realizing the topological Lifshitz
universality class, it suffices to consider only the simplest case of \(m=1\)
for which the parafermions reduce to Majoranas.  
Following \cite{Cobanera2014}, we exploit a gauge-reducing duality transformation
\cite{Cobanera2011} to map \(H_{\sf LTC}\) to a dual Landau-ordered 
system \(H_{\sf LTC}^D\). Because we fix \(m=1\), this dual system features 
spins \(S=1/2\) placed at the sites \(\r\) of a square lattice, represented by 
Pauli matrices \(\sigma^\alpha_\r\). It is governed by the Hamiltonian 
\begin{eqnarray}
H_{\sf LTC}^D=&-&\sum_\r(h_s\sigma^x_\r+h_p\mathds{1})\\
&-&J_1\sum_{\r,\mu}\sigma^z_\r\sigma^z_{\r-\bm{e_\mu}}-J_2\sum_\r 
\sigma^z_{\r-\i}\sigma^z_{\r+\i}.\nonumber
\end{eqnarray}
The dual Hamiltonian \(H_{\sf LTC}^D\) is precisely the 
celebrated quantum ANNNI model in two space dimensions. In mean field,
\(H_{\sf LTC}^D\) is directly connected to the Landau functional of 
Eq. \eqref{k_instability} \cite{Selke1988}. Since dualities are unitary 
transformations \cite{Cobanera2011}, we obtain that our perturbed toric 
code and the ANNNI model share identical phase diagrams. In the following
we rely on the extensive knowledge of this phase diagram collected in 
Ref. \cite{Selke1988}.

To characterize the non-local (dis)orders in the quantum phase diagram as 
it pertains to the topological model \(H_{\sf LTC}\), we need to identify a 
non-local order parameter. Again, we follow the ideas of Ref. 
\cite{Cobanera2013} and obtain \cite{supplemental_material} 
the string order parameter
\begin{equation}
\Sigma_\r(d)=\langle\Omega| \prod_{j=1}^{d} U_{(\r+j\i,2)}|\Omega\rangle.
\end{equation}
In terms of \(h_s\) versus \(-J_2/J_1\),
the phase diagram splits into a phase at high \(h_s\) with exponential decay
of the string order \(\Sigma_\r(d)\) and phases at low \(h_s\) with
long-range string order. The ordered phases are split
by a phase boundary starting at \(-J_2/J_1=.5, h_s=0\) into a homogeneous
phase \(k_0=0\) at low \(-J_2/J_1\), and a modulated phase for stronger
\(-J_2\), composed of (possibly infinitely!) many modulated phases with various 
\(k_0\neq0\). The two types of string orders meet the string disordered 
phase at a topological Lifshitz point. In this way, our model Hamiltonian 
\(H_{\sf LTC}\) realizes the topological Lifshitz universality class.

In summary, in this paper we have proved that competing interactions
in topological systems can lead to commensurate and incommensurate non-local 
orders with distinct critical behaviors. There are clear directions for future 
research. On the experimental side, it may be easier to demonstrate incommensurate 
non-local orders in cold atoms \cite{endres11} than in mesoscopic arrays, 
and so it would be interesting to investigate models presenting 
modulated phases for the string order parameter associated to the Haldane phase 
of \(S=1\) spin chains. On the theoretical side, it is possible that the topic 
of  modulated topological quantum orders opens an area of research significantly 
wider in scope than its Landau counterpart.
To ascertain whether this is the case it would help to characterize 
the interplay  between modulated orders and gauge fields. 
A natural, concrete starting point would be to investigate, in terms of the 
Fredenhagen-Marcu string order parameter recently rederived from dualities 
\cite{Cobanera2013}, the phase diagram of a \(\mathds{Z}_{2}\) Higgs 
model with the matter field controlled by the ANNNI model Hamiltonian.

{\it Acknowledgements.} We thank B. van Heck, 
Y. Nakata and L. Vanderstraeten for useful discussions. 
AM was supported by the ERC grants QFTCMPS and SIQS, 
and by the cluster of excellence EXC 201 Quantum Engineering
and Space-Time Research. EC was supported by the Dutch Science 
Foundation NWO/FOM and an ERC Advanced Investigator grant. 
MB acknowledges support from the German Excellence Initiative 
via the Nanosystems Initiative Munich and the EU grant SIQS.

\appendix

\section{Supplemental Material}

{\it Duality transformations.---}
We report here the duality transformations mentioned in the text, 
following closely the techniques introduced in 
Refs. \cite{Cobanera2011,clock,Cobanera2013}.

For the Hamiltonian \(H_{\sf eff}\), the duality 
transformation in question is the unitary transformation \(\mathcal{U}_{\sf d}\)  
induced by the mapping of interactions
\begin{equation}
\Gamma_i^\dagger \Delta_i\mapsto \Delta_i^\dagger \Gamma_{i+1},\quad
\Delta_i^\dagger \Gamma_{i+1}\mapsto \Gamma_{i+1}^\dagger \Delta_{i+1}\quad
(i=1,\dots,L). 
\end{equation}
The isospectral dual Hamiltonian \(H_{\sf eff}^{D}=\mathcal{U}_{\sf d}H_{\sf eff}
\mathcal{U}_{\sf d}^\dagger\) reads
\begin{eqnarray}\label{ddual_pANNNI}
H_{\sf eff}^D&=&-\frac{1}{2}\sum_{i=1}^L[(E_J \Gamma_{i}^\dagger \Delta_i\\
&+&E^{(1)}_C \Delta_i^\dagger \Gamma_{i+1} 
+E_C^{(2)}\Delta_i^\dagger \Gamma_{i+1}\Delta_{i+1}^\dagger \Gamma_{i+2}+H.c.]. \nonumber
\end{eqnarray}

It is useful to rewrite \(H_{\sf eff}^D\) in terms of local degrees of freedom. 
The combinations
\begin{eqnarray}
U_i=\Gamma_i^\dagger \Delta_i,\quad V_i=\Gamma_i\prod_{m=1}^{i-1}
\Delta^\dagger_{m}\Gamma_m,
\end{eqnarray}
of parafermions define spin-like, so-called clock variables  
that commute on different sites, and otherwise satisfy
\begin{equation}
V_iU_i=e^{i\frac{\pi}{m}}U_iV_i,\quad 
U_i^{2m}=U_iU_i^\dagger = \mathds{1}= V_iV_i^\dagger=V_i^{2m}.
\end{equation}
For \(m=1\), these relations are satisfied by letting 
\(U_i\rightarrow \sigma^z_i\) and \(V_i\rightarrow \sigma^x_i\), with
\(\sigma^x_i, \sigma^z_i\) the standard Pauli matrices. Then the reciprocal
relations
\begin{eqnarray}
\Gamma_i=V_i\prod_{m=1}^{i-1}U_m,\quad \Delta_i=\Gamma_iU_i
\end{eqnarray} 
show that, for \(m=1\), \(\Gamma_i\rightarrow a_i\) and \(\Delta_i
\rightarrow -\im b_i\), with \(a_i,b_i\) standard Majorana fermions satisfying
the standard relation \(c_i=(a_i+\im b_i)/2\) to ordinary fermions.

In terms of the local clock variables \(U_i,V_i\), and up to boundary terms that we 
neglect in the following, \(H_{\sf eff}^D\) reduces to 
\begin{eqnarray}\label{dddual_pANNNI}
&&H_{ANNNC}=\\
&-&\frac{1}{2}\sum_{i}[E_J U_i+E_C^{(1)}V_i^\dagger V_{i+1}+E_C^{(2)}
V_i^\dagger V_{i+2}+H.c.].\nonumber
\end{eqnarray}
For \(E_C^{(2)}=0\), the Hamiltonian \(H_{ANNNC}\) reduces to the 
standard clock model \cite{clock}. 
For $E_C^{(2)} < 0$,  \(H_{ANNNC}\) describes a ferromagnetic clock 
model with antiferromagnetic next-nearest-neighbor interactions. 
For \(m=1\), the clock variables are just Pauli matrices and 
\(H_{ANNNC}\) becomes the the quantum descendant of the two-dimensional
classical ANNNI model \cite{Selke1988}. Hence we call \(H_{ANNNC}\)
the anisotropic next-nearest neighbor clock (ANNNC) model. 
The phases of the ANNNC model can be distinguished by the long-distance
behavior of the two-point correlator \(g_i(d)=\langle V_i^\dagger V_{i+d}\rangle\).
This observation translates into the string order parameter 
\begin{equation}
\Sigma_i(d)=\langle \Omega|\prod_{n=i}^{i-d+1} \Gamma_n^\dagger\Delta_n^{\;}
|\Omega\rangle
\end{equation} 
for \(H_{\sf eff}\), by applying the transformations just introduced to \(g_i(d)\).

\begin{figure}[t]
\includegraphics[width=.7\columnwidth]{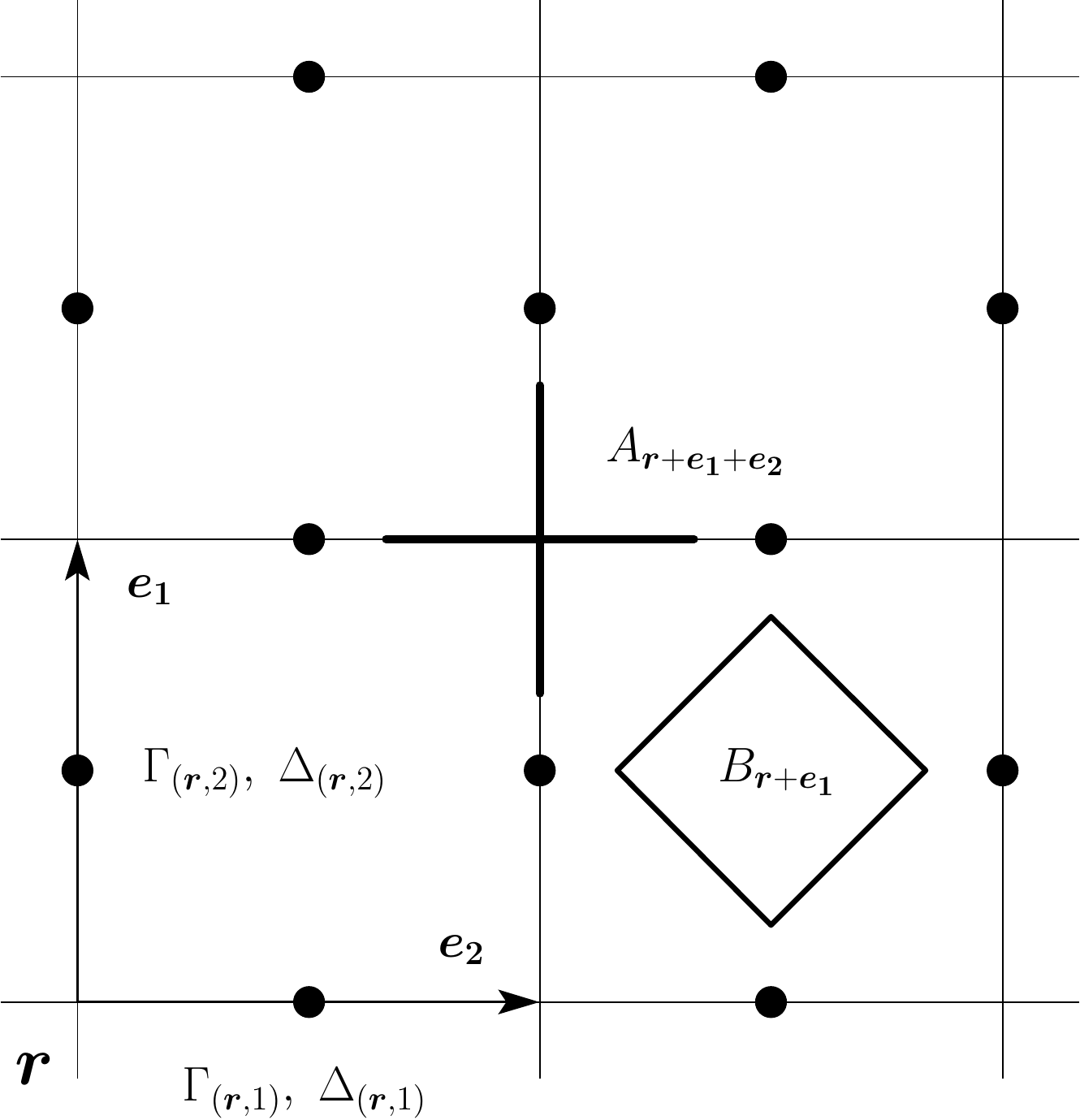}
\caption{Parafermions for the two-dimensional systems \(H_{\sf LTC}\).} 
\label{fig_sys}
\end{figure}

For the two-dimensional Hamiltonian \(H_{\sf LTC}=H_{\sf TC}+V\) (see Fig. \ref{fig_sys}
for an illustration of the notation), and \(m=1\), the duality mapping reads
\begin{eqnarray}
A_\r&\mapsto& \sigma^x_\r,\\
U_{(\r,1)}&\mapsto& \eta_{(\r,1)} \sigma^z_\r\sigma^z_{\r-\j},\\
U_{(\r,2)}&\mapsto& \eta_{(\r,2)} \sigma^z_\r\sigma^z_{\r-\i}.
\end{eqnarray}
The classical Ising variables \(\eta_{\r,\mu}=\pm 1\) are fixed in 
accordance with the relation 
\begin{equation}
B_\r\mapsto (\prod_{\mu=1}^2\eta_{(\r,\mu)}\eta_{(\r-\bm{e_\mu},\mu)})\mathds{1}
\end{equation} 
so that the dual system represents our perturbed toric code projected onto
a particular set of simultaneous eigenstates of the \(B_\r\). The gauge-invariant
sector corresponds to \(\eta_{(\r,\mu)}=1\). 
Magnetic phases that are distinguished in the ANNNI model \(H^D\) by the long-distance 
behavior of the two-point correlator
\begin{equation}
g_\r(d)=\langle \sigma^z_\r \sigma^z_{\r+d\i}\rangle
\end{equation}
are distinguished in our perturbed toric code by the string correlator
\begin{equation}
\Sigma_\r(d)=\langle \Omega|\prod_{i=1}^{d}U_{(\r+m\i,2)}|\Omega\rangle.
\end{equation}

{\it Numerical methods.---} To compute the phase diagram and wave 
vectors in Fig. \ref{fig_phase_diag} and Fig. \ref{variable_q}, we first 
obtain approximate ground states of  
$H_\text{eff}$ or $H^D_\text{eff}$ using block translation invariant MPS
\begin{align}\label{eqn:num_mps}
  |\Psi[A]\rangle = \sum_{\vec{s}=0}^p v_L^\dagger \left[\prod_{n=-\infty}^{+\infty}
    A_0^{s_{nL}} \dots A_{L-1}^{s_{(n+1)L - 1}} \right] v_R |\vec{s}\rangle,
\end{align}
where $A_k^s$ is a $D \times D$ complex matrix or parameters, 
$D$ is the bond dimension, $\vec{s} =
s_{-\infty}\dots s_{+\infty}$, and $v_L, v_R$ are boundary
vectors that do not feature in our calculations since the bulk is
completely decoupled from the infinitely distant boundaries.
To obtain a well-defined norm and expectation values, we also
require that the transfer matrix 
$E = \sum_{s_0\dots s_{L-1}} A_0^{s_0} \dots A_{L-1}^{s_{L - 1}} 
\otimes \overline{A_0^{s_0} \dots A_{L-1}^{s_{L - 1}}}$
has a unique eigenvalue of largest magnitude equal to one.

By exploiting the tangent space $\mathcal{T}_{[A]}$ \cite{tangent_plane_mps}
to the variational manifold $\mathcal{M}_D$ defined in \eqref{eqn:num_mps} 
at a given bond-dimension $D$, it is possible to compute the effective
energy gradient (imaginary time evolution), which can be used to implement
the nonlinear conjugate gradient method for minimizing the energy \cite{Milsted2013}.
The tangent plane consists of vectors $\partial_i |\Psi[A]\rangle$,
where $i$ enumerates all entries in the set of tensors $[A]$.
These methods, among others, are implemented in the open source
Python package evoMPS \cite{evoMPS}.

To obtain the approximate phase diagram of Fig. \ref{fig_phase_diag},
we fix $D$, in this case to $D=16$ or $D=24$, and compute
MPS ground states along lines in parameter space,
sweeping in both possible directions and selecting the lowest energy
state for each point. We begin with a block length of $L=1$, increasing
it if it becomes clear that the energy minimization is leading towards
a global superposition (in order to restore translation invariance),
which is indicated by the appearance of multiple eigenvalues of $E$ with
maginitude approximately equal to one. We use a variety of quantities
to locate a probable transition, in particular the first and second ground state
energy derivatives, the entanglement entropy and correlation length, the
expectation value of the order parameter, and its correlation function.
We test for criticality within a region by computing an estimate for the
CFT central charge from the scaling of the entropy and the correlation
length with the bond dimension \cite{mps_cc}.
Note that, to precisely locate and characterize a second order (or higher order)
phase transition, the bond dimension should be increased until finite entanglement
effects are no longer significant.

We estimate the wave vector of the correlation function modulation by fitting
the correlation function (or string expectation value) over twenty sites using
\begin{align}\label{eqn:corr_fit}
  f(d) = A e^{-d \lambda} \cos(k_0 d + \phi),
\end{align}
where $d$ is the distance in sites, $\lambda$ is the inverse correlation length,
$\phi$ is an offset and $k_0$ is the wave vector. We obtain an error on $k_0$
from the least squares fit result. Although the decay is approximately algebraic
(for short distances) within critical regions, this function still 
offers a good fit of the wave vector. Within a modulated critical region, 
the wave vector is also present as the phase of the second largest 
eigenvalue of $E$, the magnitude of which determines the correlation length 
\cite{tangent_plane_mps}.

For this work, we used ground state data for both $H^D_\text{eff}$ and $H_\text{eff}$,
finding the results to be consistent. $H^D_\text{eff}$ offers some numerical advantages,
possessing only nearest-neighbour interactions and having typically smaller ground 
state periodicity.

\end{document}